\documentclass[aps,pre,superscriptaddress,preprint,amsfonts,amssymb,amsmath,bibnotes,footinbib,showkeys,showpacs,titlepage]{revtex4}

\usepackage{graphicx,color}
\usepackage{ifpdf}
\usepackage[latin1]{inputenc}
\usepackage{float}

\begin{document}
\title{Intrinsic convergence properties of entropic sampling algorithms}% Force line breaks with \\
\author{Rolando Elio Belardinelli$^1$}
\author{Victor Daniel Pereyra$^1$}
\author{Ronald Dickman$^2$}
\author{Bruno Jeferson Louren\c{c}o$^2$}
\affiliation{%
$^1$Departamento de F\'{\i}sica, Instituto de F\'{\i}sica Aplicada (INFAP) - CONICET,
Universidad Nacional de San Luis, Chacabuco 917, 5700
San Luis, Argentina\\
$^2$Departamento de F\'{\i}sica and
National Institute of Science and Technology for Complex Systems,\\
ICEx, Universidade Federal de Minas Gerais, \\
C. P. 702, 30123-970 Belo Horizonte, Minas Gerais - Brazil
%\textbackslash
}%

\date{\today}% It is always \today, today,

\begin{abstract}

We study the convergence of the density of states and
thermodynamic properties in three flat-histogram
simulation methods, the Wang-Landau (WL) algorithm, the 1/t algorithm,
and tomographic sampling (TS).
In the first case the refinement parameter $f$ is rescaled ($f \to f/2$) each time
the flat-histogram condition is satisfied, in the second $f \sim 1/t$ after a suitable
initial phase, while in the third $f$ is constant ($t$ corresponds
to Monte Carlo time).
To examine the intrinsic convergence properties of these methods, free of any
complications associated with a specific model, we study a featureless
entropy landscape, such that for each allowed energy $E = 1,...,L$, there is exactly one
state, that is, $g(E) = 1$ for all $E$.
Convergence of sampling corresponds to
$g(E,t) \to $ const. as $t \to \infty$, so that the standard deviation $\sigma_g$ of
$g$ over energy values
is a measure of the overall sampling error.
Neither the WL algorithm nor TS converge: in both cases $\sigma_g$ saturates at long times.
In the 1/t algorithm, by contrast,  $\sigma_g$ decays $\propto 1/\sqrt{t}$.
Modified TS and $1/t$ procedures, in which $f \propto 1/t^\alpha$, converge
for $\alpha$ values between $0 < \alpha \leq 1$.
There are two essential facets to convergence of flat-histogram methods:
elimination of initial errors in $g(E)$, and correction of the sampling
noise accumulated during the process.
For a simple example, we demonstrate analytically, using a Langevin equation, that
both kinds of errors can be eliminated, asymptotically, if $f \sim 1/t^\alpha$
with $0 < \alpha \leq 1$.  Convergence is optimal for $\alpha = 1$.
For $\alpha \leq 0$ the sampling noise never decays, while for $\alpha > 1$
the initial error is never completely eliminated.

\end{abstract}

\pacs{02.70.Tt, 02.50.Ga, 02.50.Ng, 05.50.+q}

\maketitle

\section{INTRODUCTION}

The Wang-Landau (WL) algorithm is currently one of the most widely used Monte Carlo (MC)
simulation methods [\onlinecite{Wang01a,Wang01b,Wang02}]. Its effectiveness
is based on its simplicity and versatility in estimating the density of states (DOS) $g(E)$ with
high accuracy (here $g(E)$ represents the number of states or configurations with energy
$E$, of a given physical system).
Several recent studies propose improvements and sophisticated implementations of the WL
algorithm [\onlinecite{Day04,Zho05,Oka06,Lee06,Zho08,Chi09,Fyt10,Mal10,Than10,Bro11,tomographic}]. A
controversial point in the application of the WL and related methods is the saturation
of the error, i.e., the difference between the simulation estimate for $g(E)$ and the exact values.
In the WL algorithm the error
approaches a constant value as the number of iterations tends to infinity, as first pointed out
by Yan and de Pablo \cite{depablo03}. Several
authors [\onlinecite{Zho05,Lee06,Zho08,Earl05,Morozov07,Ca12}] have studied the accuracy,
efficiency and convergence of the WL algorithm. In particular, Zhou and Bhatt \cite{Zho05} presented
an argument for its convergence, in apparent conflict with \cite{depablo03}.

Numerical integration is an efficient test of convergence of MC methods \cite{li07}, in part
because successive points in the integration space are independent random variables.
A definitive numerical demonstration of the non-convergence of the WL algorithm can be found in
its application to numerical integration; the method generates a systematic error in the evaluation of
integrals such as the evaluation of $\pi$ \cite{Bela08}.

The principal reason for saturation of the error in the original WL algorithm is that
the refinement parameter $f$ decreases exponentially with the number of iterations, so that
in the final stages of sampling, the estimate for $g(E)$ is essentially constant.
In Refs. [\onlinecite{Bela07a,Bela07b, Bela08}] a modified
version of the WL algorithm is introduced,
in which, at long times, the refinement parameter $f$ scales as $1/t$ instead of exponentially.
This procedure, known as the {\it $1/t$ algorithm}, has been successfully applied to diverse systems
[\onlinecite{Fy08,Bin08,Bar08,Fy09,Oje09,Sin09,Mal09,Oje10,Bor10,Fy11,Ca12}].
Zhou and Su \cite{Zho08} argued that the optimal behavior of the statistical error $w(t)$
follows $w \sim t^{-\alpha}$ with $1/2 \leq \alpha \leq 1$.
Nevertheless, numerical studies show that the error in the $1/t$ algorithm
decays as $ \sqrt{1 / t}$;
to our knowledge it has not been improved upon.

Although both the WL and $1/t$ algorithms have been used to simulate many systems, several questions
remain open. For instance, does the long-time scaling of the error ($w(t) \to w_\infty >0$ in the WL
algorithm, $w(t) \sim 1/t^{1/2}$ in the $1/t$ algorithm) represent an intrinsic property of the
method, or does it depend on the physical characteristics of the problem under study?
Although the $1/t$ algorithm owes its improved convergence to the use of a refinement factor $f \propto 1/t$
at long times, it has been found that the best way to implement this algorithm is to follow the WL
prescription initially, and after some time switch to $f \propto 1/t$ [\onlinecite{Bela07a, Bela08}].
One may again ask whether this property is intrinsic to the algorithm, or if it depends on the
system being simulated.
The above questions motivate our application of the WL, $1/t$, and tomographic
algorithms to a model density of states that is completely
featureless: for each integer energy between
1 and $L$, there is one and only one state.
We may then follow the convergence
of the sampling in the simplest possible context, both numerically and analytically.
The resulting rates of convergence are
qualitatively similar to those observed in applications to physical models such as the Ising model,
which possess a nontrivial density of states, thereby suggesting that
the convergence properties are intrinsic to the sampling methods, in particular, to how
$f(t)$ is reduced at long times.
Tomographic sampling was originally implemented using, in effect, a refinement parameter $f$ that
does not change with time \cite{tomographic}.  Here we show that convergence is improved if we instead use $f \sim 1/t$
in this method as well.

The remainder of this paper is organized as follows: In Section II, after defining the sampling procedures,
the results are reported and discussed.  A simple stochastic model of the
sampling process is analyzed in Sec. III.  We present our conclusions in Section IV.

\section{Sampling procedures}

Entropic sampling MC methods, such as the WL, $1/t$ and tomographic algorithms,
typically employ a histogram $H(E)$ to refine an estimate for the entropy.
Here $H(E)$ represents the number of visits to energy $E$ during a certain iteration
of the process.
In these methods one also defines a dimensionless entropy $S(E) \equiv \ln g(E)$,
where $g(E)$ is an estimate for the number of configurations with energy $E$.
The goal of entropic sampling algorithms is to obtain precise estimates for $g(E)$
via Metropolis MC sampling, using the acceptance probability $\min \{1,g(E_j)/g(E_i)\}$,
where $E_i$ and $E_j$ denote the energies before and after a proposed transition, respectively.
Since the $g(E)$ are of course unknown {\it a priori}, the acceptance probability is evaluated
using the current estimates for the $g(E)$.
As is usual in Metropolis sampling, when a trial configuration is rejected, the existing
configuration is counted once again in the sequence.

In studies of entropic sampling, a recurrent question is whether convergence is limited by the
specific properties of the model (such as large entropy barriers), or by the method itself, due to
sampling noise.  In the present study we eliminate the former possibility by studying a
system with a trivial, featureless entropy: by definition, $g(E) = 1$ for each energy $E \in \{1,...,L\}$.
We ask to what precision the sampling methods are able to recover this result.
Since a perfect entropic sampling procedure
would yield $S(E) = Const.$, the standard deviation of $S(E)$
represents the global sampling error.  We define
\begin{eqnarray}
w(t)=\sqrt{\frac{1}{L}\sum_{E=1}^{L}[S(E,t)-\langle S(t) \rangle]^2},
\end{eqnarray}
where $\langle S(t) \rangle$ is the mean of $S$ over all states at time $t$.
We are interested in how the global error $w(t)$ scales asymptotically with time.

It is convenient to define Monte Carlo time as $t=n/L$, where $n$ is the number of attempted changes
of state, or {\it steps}.
We also define the jump length, $\lambda $, as the maximum distance accessible
at each step. Thus $\lambda=1$ means that the energy $E_n$ at step $n$ can only differ by
$\pm 1$ from $E_{n-1}$,
(under periodic boundaries), whereas $\lambda = L/2$ means that all energies are accessible at each step.
In the former case, an attempt to visit site $n$ means that the previously visited site was $n \pm 1$,
so that the acceptance probability involves a comparison of $g(E)$ at a pair of neighboring states;
in the latter case, the previously visited site can be any site in the system, with equal probability.
Thus for $\lambda = 1$, visits to states $E$ with $g(E)$ greater than their nearest neighbors
are suppressed.  This corresponds to local relaxation, since $g$ will instead
increase at one of the neighbors.  For $\lambda = L/2$ by contrast, relaxation is global,
with visiting suppressed (favored) at states with higher (lower) than average values of $g(E)$.

We now define the sampling procedures in detail.
The WL algorithm is as follows:{\sf

i) Initially, $ S (j, 0) =  H (j) = 0 $, $\forall j \in \{1,..., L\} $.
The initial refinement parameter is $ f(0) = 1$.  We define a
global stopping criterion of $f_{stop} = 10^{-8}$: sampling halts when $f(t) \leq f_{stop}$.

ii) A state $i$ is chosen at random.

iii) A state $j$ is also chosen at random between among the $2 \lambda$ states nearest $i$, that is,
from $\{i\pm 1$,...$i\pm \lambda\}$.

iv) Let
$
p = \min \{1, e^{-\Delta S}\},
\label{pac}
$
where $\Delta S=S(j,t)-S(i,t)$.
A random number, $\xi$, uniformly distributed on $[0,1]$, is generated.
If $\xi <p $, then the variables associated with state $j$ are updated so:
$S(j) = S(j) + f(t)$, $H (f) = H (f) + 1$, and the current state changes $(i \to j$);
otherwise,  $S(i, t) = S (i, t) + f (t)$ and
$ H (i) = H (i) + 1$.

v) After a fixed number of events (i.e., an MC time increment of 1000),
the flatness of the histogram $H$ is tested; if $H$ is flat$^*$,
the histogram is reset $(H(j) \to 0, \; \forall j \in \{1,..., L\})$ \cite{Flatt00}, and
the refinement parameter is reduced so: $f \to f/2$.

vi) If $ f (t) <f_{stop} $, the process stops, otherwise, a new iteration begins at item iii).}
\\

\begin{figure}
\includegraphics[scale=0.7]{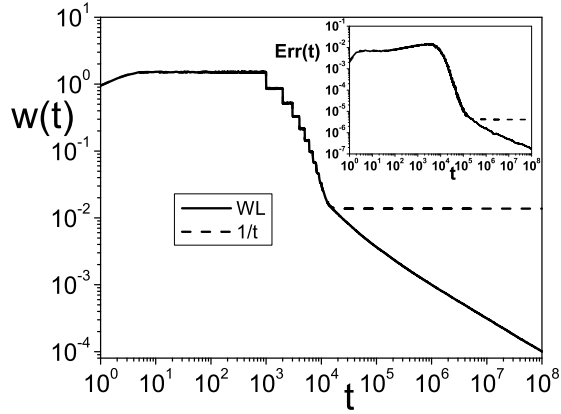}% Here is how to import EPS art
\caption{\sf {\footnotesize Global error for the uniform energy landscape.
The inset shows the error in the calculation of
$\pi$ using two-dimensional integration (see Ref.~\cite{Bela08}). Data
represent averages over $100$ independent realizations.}}
\label{globalerror}
\end{figure}

The $1/t$ algorithm \cite{Bela08} follows steps i) - v) of the above scheme, but step vi) is replaced with:

{\sf

vi) If $f \leq 1/t$ then $f \to 1/t$; thereafter, $f(t) = 1/t$ is updated at each event,
and step (vi) is not used for the remainder of the study.

vii) If $f(t)<f_{stop}$, the process stops, otherwise, a new iteration begins at item iii). }

\noindent For $\lambda = L/2$, these algorithms are similar to well known algorithms used in numerical
evaluation of multiple
integrals [\onlinecite{li07,Bela08}].

The global error, $w(t)$, found in the WL and $1/t$ procedures,
is shown in Fig.~\ref{globalerror}, for $L=256$.
The inset shows the errors corresponding to the WL and $1/t$
algorithms in the calculation of $\pi$, using two-dimensional integration (see Ref.\cite{Bela08}).
(The latter calculation uses
$I_\pi = 4 \int^{1}_0 \sqrt{1-x^2}\,\mathrm{d}x \approx   4 \sum_{i=0}^L g(y_i) y_i$ for $x$ and $y$ in the
first quadrant.
The system size is $L=256$ and $dy=1/L$.)
The time dependence of the global error is very similar in the two cases, suggesting that the convergence exhibits general features.
We can distinguish three stages in the evolution: i) a short-time regime, from the beginning to the first plateau;
ii) an intermediate regime, in which $w(t)$
diminishes $\propto \sqrt{f(t)}$ in both algorithms \cite{Zho05}; and iii) in the WL case, a long-time regime,
in which $ w (t) $ saturates.
Regime (iii) is not observed using the  $ 1 / t $ algorithm.

\begin{figure}
\includegraphics[scale=0.7]{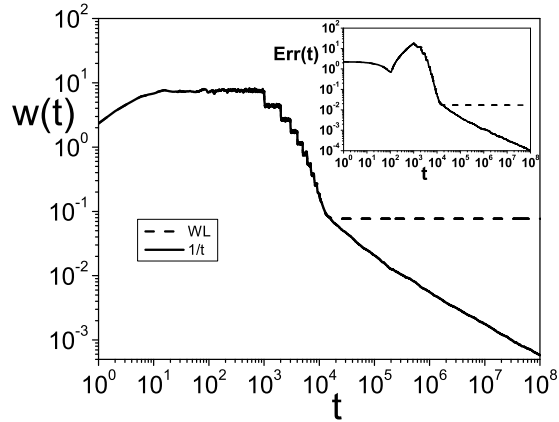}% Here is how to import EPS art
\caption{\sf {\footnotesize Global error for  $\lambda=2$.
Inset: error in the WL and $1/t$ algorithms for the 2D Ising model (see Ref.\cite{Bela07a}).
Data are the averages over $100$ independent realizations.} }
\label{ising}
\end{figure}

The general features of convergence identified above also appear in simulations of the two-dimensional
Ising model, when we compare the simulation estimates for $S(E)$ with the exact result for $L=16$
\cite{Beale96} (see Fig.~\ref{ising}).  This adds further support to the idea that the general convergence properties
of entropic sampling algorithms are model-independent.

\begin{figure}
\includegraphics[scale=0.8]{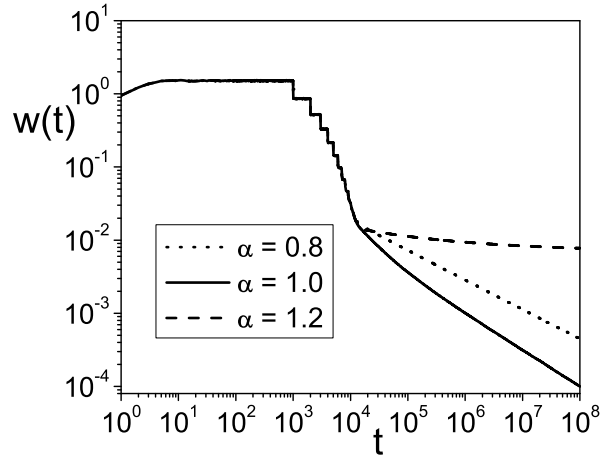}% Here is how to import EPS art
\caption{\sf {\footnotesize Global error $w(t)$ for
$f(t)=t^{-\alpha}$, with $\alpha=0.8$, $1.0$, and $1.2$. The data represent averages over
$100$ independent realizations.}}
\label{alpha}
\end{figure}

We also examine the error in a modified $1/t$ algorithm using $f(t)=t^{-\alpha}$ at long times.
In Fig.~\ref{alpha} we plot $w(t)$
for $\alpha = 0.8$, 1 and 1.2;
for $\alpha \le 1$, $w(t)\propto \sqrt{f(t)}$, whereas for
$\alpha>1$, the global error saturates at long times.  Convergence appears to be optimal for $\alpha = 1$.

\begin{figure}
\includegraphics[scale=0.8]{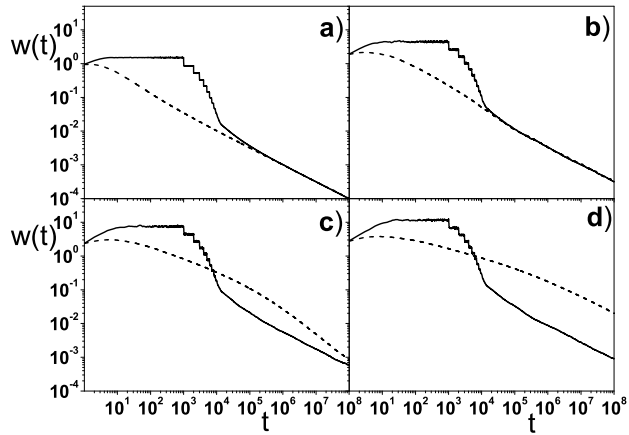}% Here is how to import EPS art
\caption{\sf {\footnotesize Global error $w(t)$ versus time for
a) $\lambda = L/2 $; b) $\lambda = 4 $; c) $\lambda = 2$; d) $\lambda = 1$,
with $L=256$.
Dashed lines: pure $1/t$ algorithm; solid lines: modified algorithm.
The data represent averages over $100$ independent realizations.}}
\label{lambda}
\end{figure}

As noted above, the $1/t$ algorithm is implemented using $f \to f/2$ (i.e., the WL rescaling) at each
iteration, {\it until $f \leq 1/t$}, at which point one switches to $f = 1/t$.  Numerical studies show that
if we take
$f(t)=1/t$ from the very beginning (the ``pure" $1/t$ algorithm), convergence is slower,
and the error may saturate in some cases.
It is nevertheless of interest to know if this is a general property of the $1/t$ algorithm, or if
it depends on the manner in which states are sampled.
The global error $w(t)$ is plotted as a function of time, for several values of $\lambda$, in
Fig.~\ref{lambda}; the full lines correspond to the (standard) hybrid $1/t$ algorithm
(i.e., initially following the WL prescription),
while the dashed lines represent the pure $1/t$ algorithm [$f(t)=1/t$ for all times $t\geq 1$].
Figure~4a shows results for $\lambda=L/2$, that is, complete independence between subsequent pairs of states
$i$ and $j$ (global relaxation).  In this case the error decays more rapidly in the pure $1/t$ algorithm
than in the hybrid algorithm.  When the new state $j$ is restricted to the neighborhood of $i$ (local relaxation),
however, the hybrid algorithm converges more rapidly, as seen in Figs.~4b-4d, which correspond to
$\lambda=4$, 2 and 1, respectively.
The difference in convergence rates between the pure and modified $1/t$
algorithms, in the local and global relaxation cases, likely reflects the fact that the modified algorithm
employs a flatness criterion, albeit only in the initial phase of sampling.
This criterion has the effect of suppressing
long-wavelength sampling fluctuations, which arise more readily under local as opposed to global relaxation.
We note that faster convergence of the hybrid $1/t$ algorithm is also found in studies of the two-dimensional
Ising model, and in the numerical evaluation of $\pi$.

\begin{figure}
\includegraphics[scale=0.4]{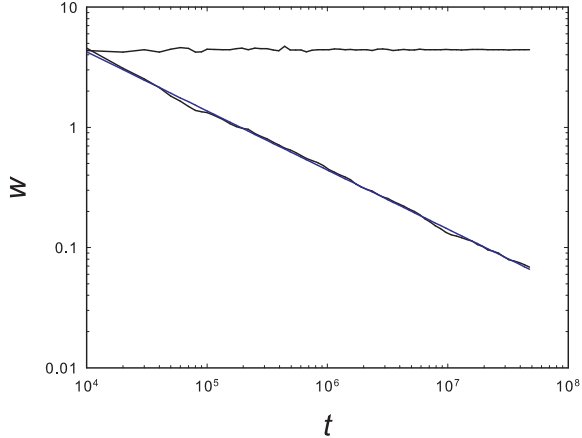}% Here is how to import EPS art
\vspace{-7em}

\caption{\label{wts} {\sf {\footnotesize (Color online) Global error $w$
versus sampling time $t$ for tomographic sampling
using $\alpha = 0$ (upper line) and $\alpha = 1$ (lower).  The straight line
fit to the data for $\alpha = 1$ has slope -0.49.  Data represent averages over 50
independent realizations of the procedure.}}}
\end{figure}

We turn now to
a related entropic simulation method, tomographic sampling \cite{tomographic}.  Here
configurations are sampled using the acceptance probability $p$ of item iv) above during
each iteration.  An iteration consists of $N_i$ studies beginning from different initial
configurations, with each study employing a large (but fixed)
number $N_s$, of MC steps.  At the end of each iteration, the estimate for $g(E)$ is updated using
$g(E,n) = [H(E)/\overline{H}]g(E,n!-!1)$, where the second argument of $g$ denotes the iteration.
($\overline{H}$ is the
average of the histogram $H(E)$ over all energies visited; $1/\overline{H}$ is used as a convenient
normalization factor.)  We applied this
procedure to the problem defined above, using 5000 iterations, each using ten initial states, uniformly
spaced over a ring of $L=256$ sites; for each initial configuration we generate a random walk
of $1000 \, L$ steps in state space.  Initially $g(E)=1, \; \forall E \in \{1,...,L\}$.
Following each update of the $g(E)$ we calculate the standard
deviation, $w$, of $g$ over the $L$ states.  Figure \ref{wts} shows that for the
procedure described above, which corresponds to $\alpha = 0$, the error $w(t)$ is essentially constant.
Modifying the procedure so that at the $n$-th iteration, $g(E,n) = [H(E)/\overline{H}]^{1/n}g(E,n)$
(corresponding to $\alpha=1$),
we observe convergence, again with $w(t) \sim t^{-1/2}$.  (Time is measured in units of lattice updates,
so that each iteration corresponds to a time increment of $10^4$.)

\subsection{Convergence of thermodynamic properties in tomographic sampling}

While the preceding discussion concerns a rather idealized situation in which the number of configurations
is the same for each energy, the scaling of the refinement factor also affects the convergence of
tomographic sampling in practical applications.  In this method, as noted above, the estimate for
$g(E)$ is updated at each iteration, with an iteration consisting of a set of studies using a large number of
updates, starting from diverse initial configurations that span the range of energies.
In \cite{tomographic} the method
was applied to the Ising model in two and three dimensions, yielding good results for critical exponents.
The studies used a series of five iterations, each involving $10^7$ lattice updates for each of ten initial
configurations, with the estimate for $g(E)$ updated using the same factor following each iteration,
corresponding to the case $\alpha=0$ discussed above.

Here we examine the quality of the results, as characterized by the statistical uncertainty of the estimates
over a set of twenty independent runs, each consisting of five iterations as described above,
in tomographic sampling of the Ising model on a square lattice of size $L=20$.
%(All results agree to within uncertainty.)
The quantities analyzed are two estimates for the critical temperature,
$T_C$ and $T_\chi$ (the temperatures at which the specific heat and the magnetic susceptibility take their
maximum values, respectively), and the maximum values, $C_{max}$ and $\chi_{max}$,
of the specific heat and susceptibility.
Using $\alpha = 0$
the relative uncertainties in $T_C$ and $T_\chi$ are both $4 \times 10^{-5}$, while for $\alpha = 1$ they fall to
$2 \times 10^{-5}$.  Similarly, the relative uncertainties in the maximum values are $\sim 3 \times 10^{-4}$
for $\alpha=0$, compared with $2 \times 10^{-4}$ for $\alpha = 1$.  Interestingly, essentially the same reduction in
uncertainties is found using $\alpha = 1/2$.  Further studies using ten iterations reveal that, for the values
of $\alpha$ considered here, there are no significant changes in the mean values or associated uncertainties after
the fourth or fifth iteration.  Although this may seem surprising given the results for the featureless entropy
landscape reported above, which show faster
convergence for $\alpha = 1$, and continued reduction of statistical errors with increasing numbers of iterations,
it should be recalled that the distribution of configurations in the Ising model is considerably more
complex than that of the uniform landscape, as is the question of the connectivity
of the configuration space via single spin flip moves, as used here and in \cite{tomographic}.

\begin{figure}
\includegraphics[scale=0.8]{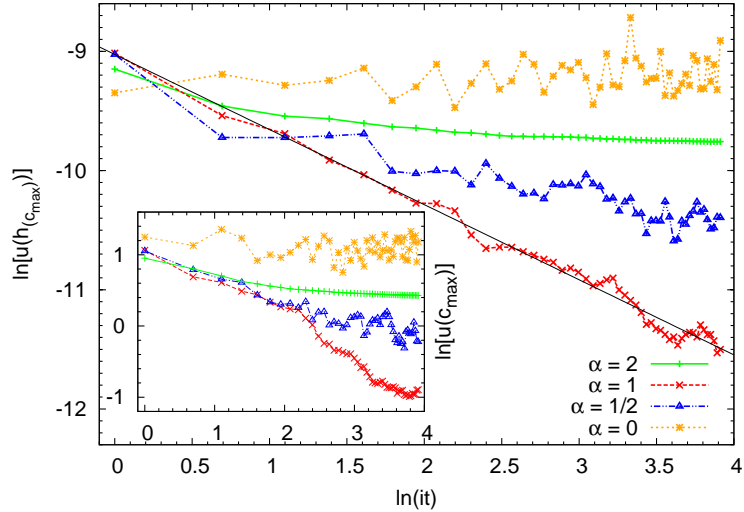}% Here is how to import EPS art
%\vspace{-7em}

\caption{\label{varhcm} {\sf {\footnotesize Antiferromagnetic Ising model in two dimensions: variance of the tomographic sampling estimate
for the external field $h$ associated with maximum specific heat versus iteration number,
for parameters as specified in text. The slope of the straight
line is -0.63.  Inset: variance of estimate for the maximum specific heat.}}}
\end{figure}

Tomographic sampling (with $\alpha=0$) was applied to the two-dimensional Ising antiferromagnet in
an external magnetic field in \cite{tomoafm}.  In this case, the sampling is bivariate, i.e., we
estimate the number of configurations with given energy {\it and} magnetization; bivariate sampling
presents a greater challenge for entropic methods than single variable sampling.
Here, we compare results using $\alpha = 0$, 1/2, 1, and 2.  In Fig.~\ref{varhcm}, for system size $L=20$,
the variance of
tomographic sampling estimates for the external field $h$ maximizing the specific heat
(at fixed temperature $T=0.2$),
is plotted versus iteration number.
(Each iteration consists of 10 studies starting from different
initial conditions, with each study comprising $10^7$ lattice updates; the variance is calculated over a sample
of 30 independent realizations.)  It is again evident that the variance saturates for $\alpha=0$, and $\alpha=2$,
and decays most rapidly (roughly $\propto t^{-0.63}$) for $\alpha=1$.  The inset shows that the maximum specific heat
behaves in an analogous manner.  In this study, employing a sequence of 50 iterations, using $\alpha=1$ instead
of $\alpha=0$ reduces the final uncertainty by an order of magnitude.  Despite this significant improvement, we
note that uncertainties in the field value that maximizes the {\it antiferromagnetic susceptibility}
$\chi$ are essentially independent of $\alpha$.  This may be because, different from the
specific heat, calculating $\chi$ requires using
``microcanonical" (fixed energy and magnetization) averages of the antiferromagnetic order parameter and
its second moment.

\section{CONVERGENCE: ANALYSIS OF A SIMPLE MODEL}

In this section we analyze convergence in a highly simplified example which nevertheless appears to capture the
essential features of entropic sampling.  The system of consists of two energy levels, each having a single state.
Let $g(1,n)$ and $g(2,n)$ denote the estimates for the numbers of states at the $n$-th iteration of the
sampling process, so that convergence corresponds to $|g(1,n) - g(2,n)| \to 0$ as $n \to \infty$.
At each iteration we generate a set of $N$ independent binary random variables $\sigma_j \in \{1,2\}$, such
that Prob$[\sigma_j = 1] = p(n)$ and Prob$[\sigma_j = 2] = q(n) = 1-p(n)$.  The number $H(n)$ of 1s in the
set $\{\sigma_1,...,\sigma_N\}$ follows the binomial distribution,

\begin{equation}
\mbox{Prob}[H(n) = m] = \binom{N}{m} [p(n)]^m [q(n)]^{N-m}
\label{binom}
\end{equation}
\vspace{.1em}

\noindent For $N \gg 1$, $H(n)$ is well approximated by a Gaussian random variable with mean $Np(n)$
and variance $Np(n)q(n)$, so that $h(n) \equiv H(n)/\overline{H} = 2 H(n)/N \simeq 2p(n) + \xi(n)$, where $\xi(n)$ is a
Gaussian random variable with mean zero and variance $4 p(n)q(n)/N \equiv \Gamma(n)$.

In entropic simulations one wishes to sample the probability distribution $P({\cal C}) = 1/g[E({\cal C})]$,
where $g(E)$ is the number of configurations with energy $E$, and $E({\cal C})$ is the energy of
configuration ${\cal C}$.  Using the current estimate for the number of configurations, we
therefore take
\begin{equation}
p(n) = \frac{g(2,n\!-\!1)}{g(1,n\!-\!1) + g(2,n\!-\!1)} .
\label{pofn}
\end{equation}
\vspace{.1em}

At each iteration $n \geq 1$ we generate revised estimates of $g(1)$ and $g(2)$ via:
\begin{equation}
g(1,n) = h(n)^{1/n^\alpha} g(1,n\!-\!1) \;\;\;\; \mbox{and}
\;\;\;\; g(2,n) = [2- h(n)]^{1/n^\alpha} g(2,n\!-\!1)
\label{update}
\end{equation}
%\vspace{.1em}

\noindent Suppose $g (1,n) < g(2,n)$. Then at the next iteration, $p>1/2$, tending to equalize
$g(1,n\!+\!1)$ and $g(2,n\!+\!1)$.
This procedure tends, on average, to reduce $|g(1) - g(2)|$.
In a given realization of the process, however,
the sampling noise $\xi(n)$ creates a fresh imbalance at each iteration, which must then be corrected
at subsequent iterations for the process to converge.

To examine the asymptotic convergence,
suppose that at step $n_0$, after various iterations, the difference $|g(1,n_0) - g(2,n_0)| \ll 1$.
Let $g(1,n_0) = 1 + u(n_0)$ and $g(2,n_0) = 1-u(n_0)$, so that
$p(n_0+1) = 1 - u(n_0)/2$.  At the next iteration we have

\begin{equation}
g(1,n_0+1) \equiv 1 + u(n_0+1) = [1 - u(n_0) + \xi(n_0) ]^{1/n^\alpha} [1 + u(n_0)]
\label{g1no1}
\end{equation}
Then to first order in the small quantities $\xi(n_0)$ and $u(n_0)$ we have

\begin{equation}
u(n_0+1) - u(n_0) = -\frac{u(n_0)}{n_0^\alpha} + \frac{\xi(n_0)}{n_0^\alpha}
\end{equation}

\noindent Associating a time interval $dt$ with each iteration, and taking a continuous-time limit,
$u(n) \to u(t)$, we obtain a Langevin equation

\begin{equation}
\frac{du}{dt} = - \frac{u}{t^\alpha} + \frac{\xi(t)}{t^\alpha} \,,
\label{langevinL}
\end{equation}
for $t>t_0$.  Here $\xi(t)$ is a white noise with intensity
$\Gamma=  1/N$.  Thus we have a linear Langevin equation in which both
the force $-u$ and the noise are scaled by $t^{-\alpha}$.  There are four cases
of interest.
\vspace{1em}

\noindent {\bf I.} {\boldmath $\alpha = 0$.}  In this case Eq.~(\ref{langevinL}) is just the Langevin equation for
the velocity of a Brownian particle in one dimension.  Under the deterministic evolution ($\xi = 0$), $u$
decays to zero exponentially,
but due to the noise we have var$[u] \to \Gamma/2  $ as $t \to \infty$, and there is no convergence.
\vspace{1em}

\noindent {\bf II.} {\boldmath $ 0 < \alpha < 1$.}  Without noise, the error decays as a stretched exponential,
\\
$u(t) \propto \exp[-t^{1\!-\!\alpha}/(1\!-\!\alpha)]$.  The noise causes the variance of $u$ to follow,

\begin{equation}
\mbox{var}[u(t)] = \Gamma \exp [-\mbox{\footnotesize ($\!\frac{2}{1\!-\!\alpha} $)} \,t^{1-\alpha}]
                   \int_{t_0}^t \exp [\mbox{\footnotesize ($\!\frac{2}{1\!-\!\alpha} $)} \,s^{1-\alpha}]
                   \frac{ds}{s^{2\alpha}}
\label{varu2}
\end{equation}

\noindent Denoting the integral by $I(t)$ and integrating by parts we have,

\begin{equation}
I(t) = \frac{1}{2s^\alpha} \exp [\mbox{\footnotesize ($\!\frac{2}{1\!-\!\alpha} $)} \,s^{1-\alpha}]
\mbox{\Large $|$}_{t_0}^t
+ \frac{\alpha}{2}  \int_{t_0}^t \exp [\mbox{\footnotesize ($\!\frac{2}{1\!-\!\alpha} $)} \,s^{1-\alpha}]
                   \frac{ds}{s^{1+\alpha}}
\label{It}
\end{equation}
\vspace{.1em}

\noindent The integrated part grows asymptotically as
$\frac{1}{2 t^\alpha} \exp [\mbox{\footnotesize ($\!\frac{2}{1\!-\!\alpha} $)} \,t^{1-\alpha}] \equiv F(t)$,
whereas the integral in the second term in Eq.~(\ref{It}) is smaller than $I(t)$, since $\alpha < 1$.
Thus, as $t \to \infty$,
we have $F(t) \leq I(t) \leq \frac{2}{2-\alpha}F(t)$, so that var$[u(t)] \sim 1/t^\alpha$, and the rms
error of the result decays as $1/t^{\alpha/2}$.
\vspace{1em}

\noindent {\bf III.} {\boldmath $\alpha = 1$.}  Without noise, $u(t)$ decays algebraically, $u \sim 1/t$.
Including the noise, var$[u] \sim 1/t$, and the rms error decays as $1/t^{1/2}$.
\vspace{1em}

\noindent {\bf IV.} {\boldmath $ \alpha > 1$.}  The solution to the noise-free equation is
$u = \exp [\mbox{\footnotesize $\frac{1}{(\alpha \!-\!1)t^{\alpha-1}}$}] $, so that even
deterministically, there is no convergence.  The variance is

\begin{equation}
\mbox{var}[u(t)] = \Gamma \exp [\mbox{\footnotesize $\frac{2}{(\alpha \!-\!1)t^{\alpha-1}}$}]
                   \int_{t_0}^t \exp [-\mbox{\footnotesize $\frac{2}{(\alpha \!-\!1)s^{\alpha-1}}$}]
                   \frac{ds}{s^{2\alpha}} \,.
\label{varu4}
\end{equation}

\noindent Calling the integral $I'(t)$,
it is easy to show that,

\begin{equation}
I'(t) \geq \frac{\exp \left[-\mbox{\footnotesize $\frac{2}{(\alpha \!-\!1)t_0^{\alpha-1}}$} \right]} {2 \alpha -1}
\left(\frac{1}{t_0^{2 \alpha - 1}} - \frac{1}{t^{2 \alpha - 1}} \right)
\label{It4}
\end{equation}

\noindent which implies that var$[u(t)]$ remains nonzero as $t \to \infty$.
\vspace{1em}

Summarizing, there is convergence only in cases II and III, that is, for $0 < \alpha \leq 1$, leading to
var$[u(t)] \sim t^{-\alpha/2}$.  Increasing $\alpha$
within this interval, the noise-induced error decays more rapidly, but the deterministic convergence is slower.
The WL algorithm corresponds, in this example, to Eq.~(\ref{langevinL})
with the factor $1/t^\alpha$ replaced by one which decays
exponentially, and so, as in case IV above, there is no convergence.

The foregoing analysis is readily generalized to a set of $L$ states, each represented by a single configuration.
Assuming that the random walk over configuration space is able to sample all states uniformly,
the decay of each deviation $u(k) = g(k)-1$ follows, in the linearized analysis, a Langevin equation (\ref{langevinL}),
but with noise intensity $\Gamma = L^2/N$.

As a further generalization, suppose there are two classes containing different numbers of configurations, $r_1$
and $r_2$.  The target values are $g(i) = 2r_i/M \equiv \overline{g}(i)$, where $M=r_1 + r_2$
and we have maintained the normalization
$g(1)+g(2) = 2$.  Suppose, as before, that after a number $n_0$ of iterations, the estimates are close to the
target values, so that $g(i,n_0) = \overline{g}(i) + u(i,n_0)$ with $|u(i,n_0)| \ll 1$.
Repeating the linear analysis used above, we verify that
the $u(i)$ satisfy Eq.~(\ref{langevinL}) with noise intensity $\overline{g}(i)^2/N$.  The asymptotic
convergence properties are therefore as found above for the case $r_1=r_2=1$.

\section{CONCLUSIONS}

We study convergence properties of flat histogram entropic sampling methods, specifically the Wang-Landau
and $1/t$ algorithms,
and tomographic sampling.
There appear to be two difficulties to achieving precise results in entropic sampling.
The first is associated with visiting the entire configuration space uniformly.  For small systems
this poses no problem, but it rapidly becomes more difficult, demanding ever longer simulation
times, as the system size increases.  The second difficulty, which is the main focus of present work,
concerns eliminating the effects of initial errors, and, crucially, sampling noise accumulated over the iterations.
Numerical studies of a model with a flat entropy landscape
yield the conclusion that asymptotically, the error decays most rapidly in the $1/t$-algorithm,
with the global error decaying $\propto 1/t^{1/2}$.
A modified tomographic sampling scheme, in which corrections to the estimates for the number of states $g(E)$
are weighed $\propto 1/t$,
yields a similar convergence.  We present evidence that
this behavior is general, independent of the physical model under study.
The convergence of microcanonical averages, however, remains as a topic for future study.

An important conclusion is that the error converges to a nonzero value after a certain number of
iterations in the WL algorithm, and does not decrease in subsequent iterations.  This point,
which was noted in \cite{Ca12}, can be understood on the basis of a simple stochastic model,
which also explains the observed convergence behaviors of the error in $1/t^\alpha$ schemes.

Our results would appear to conflict with those of \cite{Zho05}, which reports convergence of the WL
algorithm.  They can be reconciled by observing that the latter work demonstrates convergence
as the sample size $N \to \infty$, whereas here we consider iterations with a fixed, finite $N$, and analyze how
the dependence of the refinement factor $f(t)$ on the time or iteration number $t$ affects convergence
as $t \to \infty$.
\vspace{2em}

\noindent {\bf Acknowledgments}
\vspace{1em}

This work is partially supported by the CONICET (Argentina) and CNPq (Brazil).


\begin{thebibliography}{99}
\bibitem{Flatt00} We use a flatness criterion such that, for each $j$, the histogram $H(j)$
is at least 80\% of $\langle H \rangle=\frac{1}{L}\sum_{i=1}^{L}H(i)$ \cite{Wang01b}.
\end{thebibliography}
\end{document}